\documentstyle{article}
\Roman{section}

\begin{document}
\begin{titlepage}
\centerline{\large \bf A POSSIBLE EXPLANATION WHY $\tau_{B^{\pm}}\sim
\tau_{B^0}$ BUT $\tau_{D^{\pm}}\sim 2\tau_{D^0}$}

\vspace{2.5cm}

\centerline{ Hui-Shi Dong$^{2}$, Xin-Heng Guo$^{3,4}$,
Xue-Qian Li$^{1,2}$,
and Rui Zhang$^{2}$}
\vspace{1cm}
{\small
\begin{center}

1.   CCAST  (World Laboratory)
P.O.Box 8730 Beijing 100080, China.
\vspace{8pt}

2. Department of Physics, Nankai University,
Tianjin, 300071, China.
\vspace{8pt}

3. Department of Physics and Mathematical Physics and \\
Special Research 
Center for Subatomic Structure of Matter, University of
Adelaide, SA 5005, Australia.
\vspace{8pt}

4. Institute of High Energy Physics, Academia Sinica,
Beijing 100039, China.

\end{center}
}
\vspace{2cm}

\begin{center}
\begin{minipage}{12cm}
{\bf Abstract}
\vspace{1cm}

Data show that $\tau_{B^{\pm}}\sim\tau_{B^0}$, but $\tau_{D^{\pm}}\sim
2\tau_{D^0}$. The naive interpretation which attributes $\tau_{D^{\pm}}\sim
2\tau_{D^0}$ to a destructive interference between two quark diagrams for
$D^{\pm}$ decays, definitely fails in the B-case. We 
investigate Close and Lipkin's suggestion that the phases for producing
radially excited states $\psi_{2s}$  in the decay products of B-mesons
can possess an opposite sign to the integrals for $\psi_{1s}$ decay products.
Their contributions can partially compensate each other  to result in
$\tau_{B^{\pm}}\sim\tau_{B^0}$. Since D-mesons are much lighter than B-mesons,
such possibilities do not exist in D-decays.
\end{minipage}
\end{center}

\vspace{1cm}

\end{titlepage}

\baselineskip 22pt

\centerline{\large\bf A Possible Explanation Why $\tau_{B^{\pm}}\sim
\tau_{B^0}$ But $\tau_{D^{\pm}}\sim 2\tau_{D^0}$}

\vspace{1cm}

\noindent{\bf I. Introduction.}
\vspace{0.3cm}

The naive explanation for $\tau_{D^{\pm}}\sim 2\tau_{D^0}$ \cite{Data} is
that a destructive interference between two quark diagrams for $D^{\pm}$
\cite{Ruckl} reduces the strength of decay amplitudes and thereby elongates
the life of $D^{\pm}$.
More explicitly, if the lifetime of a meson is mainly determined by the
Cabibbo favored decay modes, for $D^+$ there is only one topology
$D^+\rightarrow \bar K^0 M^+$, (where $M$ generically refers to
$\pi, \rho$ etc and $K$ to strange mesons)
whereas for $D^0$ there are two channels $D^0\rightarrow \bar K^0 M^0$
and $D^0\rightarrow K^- M^+$. For the $D^+$ decays, the two quark
diagrams shown in Fig.1 (a) and (b) interfere, while for $D^0$, the
two diagrams (c) and (d) correspond to two different modes, so do not
interfere. For the B decays, similar diagrams exist and there could be also
destructive interference in $B^-$ decays. However, the experimental data
show that $\tau_{B^{\pm}}\sim \tau_{B^0}$ \cite{Data}.

The explanation for the lifetime differences in D and B cases invloves
nonperturbative QCD phenomena. 
Actually some authors
\cite{Buras}\cite{bigi} proposed the so-called Pauli Interfernce (PI) 
mechanism as a correction to ``pure''  spectator mechanism for taking 
into account of the light degrees of freedom.
The PI effects only exist in 
$D^{\pm}$ and $B^{\pm}$ decays but not in $D^0$ and $B^0$ decays.
Based on QCD, Bigi et al. \cite{bigi} introduced a virtual gluon 
so that one of the quark produced by the weak
decay of the heavy quark interferes with the spectator quark.
In this mechanism,
the PI term modifies the ``pure'' spectator diagram and it is found that such 
interference is destructive and is proportional to $\Gamma_0 /m_{Q}^{2}$ 
(Q=b or c).
This mechanism partly explains why $\tau_{D^{\pm}}\sim 2\tau_{D^0}$ 
and $\tau_{B^{\pm}}\sim \tau_{B^0}$.

In the present work we try to investigate the lifetime differences
in another way which is based on the idea of Close and Lipkin.
Recently Close and Lipkin\cite{cl97} have analysed the data on low lying
exclusive quasi-two body final states in both $D$ and $B$ decays. They noted
that in $D$ decays the sign of interference in exclusive channels is
still amiguous while in $B$ decays there is a clear and uniform tendency
towards constructive interference between the color favoured and
colour suppressed exclusive channels where all final state mesons have
nodeless wave functions. They noted that in $B$ decays, in order that 
$\tau_{B^{\pm}}\sim \tau_{B^0}$ \cite{Data}, this interference must
be compensated in as yet unmeasured channels. They suggested that
the sign of interference may be changed in channels where excited states
of the decay products, whose wavefucntions contain nodes, are involved. 
It is the motivation of the present paper to include the contributions
from the excited states of the B decay products so that constructive
interference is obtained in $B^{\pm}$ decays. Such excited states only
excist in B decays but not in D decays because of phase space requirement.
It will be shown that in our model the lifetime differences in B and D
mesons can also be explained.


The effective Hamiltonian of non-leptonic decays in the D-case
\cite{Gaillard}\cite{Gilman} is
\begin{equation}
\label{hal}
H_{eff}={G_F\over\sqrt 2}V^*_{cs}V_{ud}
[c_1\bar s\gamma_{\mu}(1-\gamma_5)c\bar u
\gamma^{\mu}(1-\gamma_5)d+c_2\bar s\gamma_{\mu}(1-\gamma_5)d\bar u
\gamma_{\mu}(1-\gamma_5)c],
\end{equation}
where $c_1={c_++c_-\over 2}$ and $c_2={c_+-c_-\over 2}$.
By the renormalization group equation (RGE) we have
\begin{equation}
\label{coef}
c_-=({\alpha_s(m_c^2)\over\alpha_s(m_b^2)})^{12/25}
({\alpha_s(m_b^2)\over\alpha_s(M_W^2)})^{12/23}; \;\;\;\;\; c_+={1\over
\sqrt{c_-}}.
\end{equation}
With the Fiertz transformation, the coefficients $c_1$ and $c_2$ in
eq.(\ref{hal}) should be replaced by $a_1$ and $a_2$ with
\begin{equation}
\label{fac}
a_1=c_1+\xi c_2, \;\;\; {\rm and}\;\;\; a_2=c_2+\xi c_1,
\end{equation}
where $\xi$ is $1/N_c$ if the factorization assumption holds perfectly, otherwise
$\xi=(1+\delta)/N_c$ where $\delta$ denotes a color-octet contribution
proportional to $<\lambda^a\lambda^a>$ \cite{Bauer}\cite{Buras}\cite{Li}.
Recently, Blok and Shifman gave a more theoretical estimation \cite{Blok},
but they also pointed out that the obtained value is not accurate for
practical calculations. Generally, $\delta$ is a negative number ranged
between 0 to $-1$, so that $\xi$ takes values between 0 to $1/N_c$. Later,
in our numerical calculations we will take $\delta$ as 0,
$-0.5$ and $-1$ respectively.

For the B-case, we have a similar Hamiltonian as eq.(\ref{hal})
\begin{equation}
\label{halb}
H_{eff}={G_F\over\sqrt 2}V_{cb}V^*_{ud}
[c^{(B)}_1\bar c\gamma_{\mu}(1-\gamma_5)b\bar d
\gamma^{\mu}(1-\gamma_5)u+c^{(B)}_2\bar c\gamma_{\mu}(1-\gamma_5)u\bar d
\gamma_{\mu}(1-\gamma_5)b]
\end{equation}
and coefficients $c_1^{(B)}$ and $c_2^{(B)}$
\begin{equation}
\label{coefb}
c^{(B)}_-=({\alpha_s(m_b^2)\over\alpha_s(M_W^2)})^{12/23};
\;\;\;\;\; c^{(B)}_+={1\over\sqrt{c^{B}_-}},
\end{equation}
whereas $a_1^{(B)}$, $a_2^{(B)}$ have similar forms
in analog to that for the charm case.

It is noted that in the case of D meson decays  $a_1$ is positive and 
and $a_2$ is negative. From the data of D-physics the value of $a_2$ is
about $-0.5$ \cite{a2} . In $D^+$ decays, $a_1$ term corresponds
to the external W-emission while $a_2$ to the internal W-emission, naturally
a destructive interference would occur between the two quark diagrams.

In the following we will express the corresponding transition amplitudes as
$A_1$ and $A_2$ which are proportional to $a_1$ and $a_2$ respectively, thus
$A_1=\kappa_1a_1$ and $A_2=\kappa_2a_2$ where $\kappa_1$ and $\kappa_2$ are
the hadronic transition matrix elements.

Then we have the amplitude square as,
\begin{equation}
|<\bar K^0\pi^+|H_{eff}|D^+>|^2=|A_1|^2+|A_2|^2+2Re(A_1A^*_2).
\end{equation}
Beside a common phase factor such as the Cabibbo-Kabayashi-Maskawa phase,
both $A_1$ and $A_2$ are real.
Thus if $A_1\cdot A_2$ is negative, this is a destructive interference. 
Otherwise we have constructive interference.

In contrast, for $D^0$ decays,
$$<K^-\pi^+|H_{eff}|D^0>\propto a_1\;\;\;\;{\rm and}\;\;\;
<\bar K^0\pi^0|H_{eff}|D^0>\propto a_2. $$
We can roughly assume
$$<K^-\pi^+|H_{eff}|D^0>\approx A_1\;\;\;\;{\rm and}\;\;\;
<\bar K^0\pi^0|H_{eff}|D^0>\approx A_2. $$

Thus if we only consider the C-K-M favored channels which dominate the
lifetime of D-mesons, we have
\begin{eqnarray}
\Gamma(D^+) &=& (|A_1|^2+|A_2|^2+2Re(A_1A_2^*))\times LIPS \\
\Gamma(D^0) &=& (|A_1|^2+|A_2|^2)\times LIPS,
\end{eqnarray}
where LIPS is the Lorentz-invariant-phase-space of the final products.
If $A_2\sim -0.26A_1$, one can numerically obtain $\Gamma(D^0)\sim
2\Gamma(D^+)$ (or $\tau_{(D^{\pm})}\sim 2\tau_{D^0}$). Of course, the
other channels (Cabibbo-suppressed) and semi-leptonic decays all contribute
to the lifetime, so this obtained number is not rigorous. However, since
the Cabibbo favored channels dominate, one can expect that
a solution for $A_1$ and $A_2$ does not deviate much from the aforementioned
value.

Taking $\alpha_s(M_Z^2)=0.118$ \cite{Data}, one can obtain a ratio of
$A_1/A_2$ for D-decays
to be roughly consistent with the required value. By our recent
knowledge, the hadronic matrix elements can be evaluated
more easily in terms of the heavy quark effective theory (HQET) \cite{Isgur}.

In the same scenario and  by eqs.(\ref{halb}), (\ref{coefb})
$a_1$ and $a_2$ are still of opposite signs in B-decays. It is a
consequence of renormalization group equation which is proved to be valid
for perturbative QCD. If so, one could expect a result similar to the D-case
that $\tau_{B^-}\sim 2\tau_{B^0}$. However, this does not coincide with
the data for B-decays.

The $B^{(\pm)}$ lifetime is very close to that of $B^0$ as $\tau_{B^{(\pm)}}
\sim(1.62\pm 0.06)\times 10^{-12}$ s and $\tau_{B^0}\sim (1.56\pm 0.06)\times
10^{-12}$ s \cite{Data}. There could be small measurement uncertainty as
$\tau_{(B^{\pm})}\sim 1.47\times 10^{-12}$ s,
$\tau_{(B^0)}\sim 1.25\times 10^{-12}$ s, by the ALEPH collaboration
\cite{ALE}\cite{ALE1}
and $\tau_{(B^{\pm})}\sim 1.72\times 10^{-12}$ s,
$\tau_{(B^0)}\sim 1.63\times 10^{-12}$ s, by the DELPHI collaboration
\cite{DEL}.

Similar quark diagrams exist in B-decays, namely there are both external and
internal W-emissions for $B^-\rightarrow D^0\pi^-$ which destructively
interfere, but for $B^0$, $B^0\rightarrow D^+\pi^-$ and $B^0\rightarrow
D^0\pi^0$ corresponding to external and internal W-emissions respectively
do not interfere. Thus if that is the case, one would wonder why
$\tau_{B^{\pm}}$ is so close to $\tau_{B^0}$.

To fit the data of B decays, one needs to take a positive value for $a_2$
\cite{a2}. This contradicts to the result of RGE which is obviously
correct by the perturbative QCD theory and  there is
no doubt of application of perturbative QCD at the $m_b$ energy region.

However, one can notice that even though $A_1,\; A_2$ are proportional to
$a_1,\; a_2$ respectively, they also possess certain factors corresponding to
the hadronic matrix elements. These hadronic matrix elements involve
some overlapping integrations of the decay parent and daughter wavefunctions.
If the integrations can contribute a negative sign, the interference
between two diagrams would turn over to be constructive and it may be
equivalent to an "effective" positive $a_2$ value.

The hadronization process is very non-perturbative and we cannot evaluate it
accurately, so that we attribute the non-perturbative effects into
the parameters of meson wavefunctions which exist in the overlapping
integration.
To evaluate such overlapping integrations, one needs to invoke some
concrete models and later we employ the non-relativistic quark model. Since
the decaying B-meson is a pseudoscalar at $\psi_{1s}$ radial ground state, if
the decay product is at $\psi_{1s}$ state, the overlapping integration would
certainly be positive, however, if the decay products can be radially
excited states $\psi_{2s}$, the integration can turn sign (see next section
for details). Because D-meson is much lighter than B-meson, it does not
seem to exist $\psi_{2s}$ states as decay products of D, but definitely
there should be $\psi_{2s}$ excited states showing up as decay
products of B-meson. This change may modify the whole picture and finally
leads to a consequence that $\tau_{B^{(\pm)}}\sim\tau_{B^0}$.
Later our numerical results will show that the involvement of the $\psi_{2s}$
decay products can indeed do the job.

In next section, we give the formulation in every detail and in Sec.III,
we present our numerical results while the last section is devoted to
conclusion and discussion.\\

\noindent{\bf II. Formulation}
\vspace{0.3cm}

(i) The transition amplitudes.

As usual, we ignore the W-exchange and annihilation diagrams because the
two fast quarks would pick up a quark-pair from vacuum and speed them up
\cite{Bauer}. Even though the factorization approach is not very reliable to
evaluate the internal W-emission diagrams, we may use a phenomenological
parameter $\delta$ to compensate it. Therefore by the vacuum saturation
\begin{eqnarray}
&& <K^-\pi^+|a_1(\bar sc)(\bar ud)+a_2(\bar sd)(\bar uc)|D^0> = \nonumber \\
&& a_1<\pi^+|(\bar ud)|0><K^-|(\bar sc)|D^0>+a_2<K^-\pi^+|(\bar sd)|0>
<0|(\bar uc)|D^0>= \nonumber \\
&& a_1f_{\pi}p_{\pi}^{\mu}<K^-|(\bar sc)|D^0>+a_2f_Dp_D^{\mu}<K^-\pi^+|
(\bar sd)|0>
\end{eqnarray}
where $(\bar qq')\equiv \bar q\gamma_{\mu}(1-\gamma_5)q'$. The
second term corresponds to an W-annihilation diagram and obviously is much
smaller than the first one as it is proportional to $f_D(m_K^2-m_{\pi}^2)$.
As argued in literatures this term is negligible and we will omit such
contributions in later calculations. Then we also have
\begin{equation}
<\bar K^0\pi^0|a_1(\bar sc)(\bar ud)+a_2(\bar sd)(\bar uc)|D^0> =
a_2f_Kp_K^{\mu}<\pi^0|(\bar uc)|D^0>,
\end{equation}
and
\begin{equation}
<\bar K^0\pi^+|a_1(\bar sc)(\bar ud)+a_2(\bar sd)(\bar uc)|D^+> =
a_1f_{\pi}p_{\pi}^{\mu}<\bar K^0|(\bar sc)|D^+>+a_2f_Kp_K^{\mu}<\pi^+|
(\bar uc)|D^+>.
\end{equation}
Instead, for $P\rightarrow PV$
\begin{eqnarray}
<K^-\rho^+|a_1(\bar sc)(\bar ud)+a_2(\bar sd)(\bar uc)|D^0> &=&
a_1f_{\rho}m_{\rho}\epsilon^{*\mu}<K^-|(\bar sc)|D^0>, \\
<K^{-*}\pi^+|a_1(\bar sc)(\bar ud)+a_2(\bar sd)(\bar uc)|D^0> &=&
a_1f_{\pi}p_{\pi}^{\mu}<K^{-*}|(\bar sc)|D^0>,
\end{eqnarray}
\begin{eqnarray}
<\bar K^{0*}\pi^0|a_1(\bar sc)(\bar ud)+a_2(\bar sd)(\bar uc)|D^0> &=&
a_2f_{K^*}m_{K^*}\epsilon^{*\mu}<\pi^0|(\bar uc)|D^0>,\\
<\bar K^0\rho^0|a_1(\bar sc)(\bar ud)+a_2(\bar sd)(\bar uc)|D^0> &=&
a_2f_Kp_K^{\mu}<\rho^0|(\bar uc)|D^0>,
\end{eqnarray}
and
\begin{eqnarray}
<\bar K^{0*}\pi^+|a_1(\bar sc)(\bar ud)+a_2(\bar sd)(\bar uc)|D^+> &=&
a_1f_{\pi}p_{\pi}^{\mu}<\bar K^{0*}|(\bar sc)|D^+>+ \nonumber \\
&& a_2f_{K^*}
\epsilon_{K^*}^{*\mu}m_{K^*}<\pi^+|(\bar uc)|D^+>,\\
<\bar K^0\rho^+|a_1(\bar sc)(\bar ud)+a_2(\bar sd)(\bar uc)|D^+> &=&
a_1f_{\rho}\epsilon_{\rho}^{*\mu}m_{\rho}<\bar K^0|(\bar sc)|D^+>+ \nonumber\\
&& a_2f_Kp_K^{\mu}<\rho^+|(\bar uc)|D^+>.
\end{eqnarray}
For $P\rightarrow VV$
\begin{eqnarray}
<\bar K^{0*}\rho^0|a_1(\bar sc)(\bar ud)+a_2(\bar sd)(\bar uc)|D^0> &=&
a_2f_{K^*}m_{K^*}\epsilon^{*\mu}<\rho^0|(\bar uc)|D^0>,\\
<K^{-*}\rho^+|a_1(\bar sc)(\bar ud)+a_2(\bar sd)(\bar uc)|D^0> &=&
a_1f_{\rho}m_{\rho}\epsilon^{*\mu}<K^{-*}|(\bar sc)|D^0>,
\end{eqnarray}
and
\begin{eqnarray}
<\bar K^{0*}\rho^+|a_1(\bar sc)(\bar ud)+a_2(\bar sd)(\bar uc)|D^+> &=&
a_1f_{\rho}\epsilon_{\rho}^{\mu}m_{\rho}<\bar K^{0*}|(\bar sc)|D^+>+
\nonumber \\
&& a_2f_{K^*}
\epsilon_{K^*}^{*\mu}m_{K^*}<\rho^+|(\bar uc)|D^+>.
\end{eqnarray}

The above formulae indicate that the external and internal W-emissions
in $D^+$ decays interfere.

For the B-case, we can have similar expressions with an effective
hamiltonian eq.(\ref{halb}) and corresponding coefficients $a_1^{B},\;
a_2^{B}$ in eq.(\ref{coefb}).\\

(ii) The matrix elements.

It is noted that in the scenario of factorization, the hadronic matrix
elements are related to a weak transition \cite{Wirbel}, 
for $P\rightarrow P$
\begin{equation}
\label{PP}
<X|j_{\mu}|I>=(P_I+P_X-{M_I^2-M_X^2\over q^2}q)_{\mu}F_1(q^2)+{M_I^2
-M_X^2\over q^2}q_{\mu}F_0(q^2),
\end{equation}
with $q\equiv P_I-P_X$ and $F_1(0)=F_0(0)$. For $P\rightarrow V$, we have
\begin{eqnarray}
\label{PV}
<X^*|j_{\mu}|I> &=& {2\over M_I+M_{X^*}}\epsilon_{\mu\nu\rho\sigma}
\epsilon^{*\nu}P_I^{\rho}P_{X^*}^{\sigma}V(q^2)+i{\epsilon^*\cdot q\over
q^2}2M_{X^*}q_{\mu}A_0(q^2)+ \nonumber \\
&& i\{\epsilon^*_{\mu}(M_I+M_{X^*})A_1(q^2)-({\epsilon^*\cdot q\over
M_I+M_{X^*}})(P_I+P_{X^*})_{\mu}A_2(q^2)- \nonumber \\
&& {\epsilon^*\cdot q\over q^2}2M_{X^*}q_{\mu}A_3(q^2)\},
\end{eqnarray}
with $A_3(0)=A_0(0)$ and here
\begin{equation}
A_3(q^2)={M_I+M_{X^*}\over 2M_{X^*}}A_1(q^2)-{M_I-M_{X^*}\over 2M_{X^*}}
A_2(q^2).
\end{equation}

So our task is to calculate the form factors. Taking the nearest pole
approximation
\begin{eqnarray}
&& F_1(q^2)\approx {h_1\over 1-q^2/M_1^2} \;\;\;\;\;\; {\rm for}\;\; P_I\rightarrow
P_X, \\
&& V(q^2)\approx {h_V\over 1-q^2/M_2^2},\;\;\; A_0(q^2)={h_{_{A_0}}\over
1-q^2/M_3^2}\;\;\;\;{\rm for}\;\; P_I\rightarrow P_{X^*},
\end{eqnarray}
where $M_1,\; M_2,\; M_3$ are masses of mesons corresponding to the
nearest poles which can be found in the data book. With this approximation,
to evaluate the form factors, one only needs to calculate the constant
parameters $h_0=h_1$, $h_V,\; h_{A_1},\; h_{A_2}$ and $h_{A_3}=h_{A_0}$,
which turn out to be the values of the form factors at the unphysical
kinematic region $q^2=0$ and we will use the non-relativistic quark model
to calculate them. Moreover, for the case of a pseudoscalar B or D meson
transiting to a vector meson, we use the helicity amplitude method
\cite{Chung} which can much simplify our calculations.

The parameters are related to an overlapping integral over the wavefunctions
of initial pseudoscalar and final pseudoscalar or vector mesons. To carry out
the integration, one needs to invoke concrete models and the most popular
one is to take the wavefunction of harmonic oscillation potential as the
orbital part. In the Bauer-Stech-Wirbel approach \cite{Wirbel} the 
following wavefunction model is employed
\begin{equation}
R_m(p_T, x)=N_m\sqrt{x(1-x)}\exp(-p_T^2/2\omega^2)\cdot \exp({-m^2\over 2a^2}
(x-{1\over 2}{m_{q_1}^2-m_{q_2}^2\over 2m^2})^2)
\end{equation}
where $N_m$ is the normalization factor while Guo and Huang\cite{Wirbel}
used the following wavefunction form in the light-cone formalism
\begin{equation}
R_m(x,k_{pert})
=A\exp(-b^2({k_{\bot}^2+m_1^2\over x_1}+{k_{\bot}^2+m_2^2\over x_2})).
\end{equation}
These wavefunctions apply in the infinite-momentum frame. Here instead, we
choose the wavefunction at the rest frame of the decaying meson \cite{Le}.
Everything in the picture is non-relativistic, but it is accurate enough
for the qualitative conclusion and we will discuss it in the final section.

Here we only list the radial wavefunctions of $\psi_{1s}$  and $\psi_{2s}
$  and the others can be found in ref.\cite{Page}
\begin{equation}
\label{wav}
\psi_{1s}=({4\beta^3\over\sqrt{\pi}})^{1/2}\exp(-{1\over 2}\beta^2r^2)\sqrt m
Y_{00}(\theta,\phi),
\end{equation}
where adding a factor $\sqrt m$ is for proper normalization,
and
\begin{equation}
\label{wav1}
\psi_{2s}=({4\beta^3\over 6\sqrt\pi})^{1/2}(3-2\beta^2r^2)\exp
(-{1\over 2}\beta^2r^2)Y_{00}(\theta,\phi)\sqrt m,
\end{equation}
where $\beta$ is the only free parameter to be fixed by data and here
$r\equiv |\vec r_1-\vec r_2|$ in the potential picture. To convert into the
momentum-space, we have
\begin{eqnarray}
&& h_1=h_0 = \nonumber \\
&& {2m_I\over m_I^2-M_X^2}\cdot \int d^3p_1\phi_X^*(\vec p_{1'})
\phi_I(\vec p_1)
({p_{1'}^3\over p^0_{1'}+m_{1'}}
+{p_1^3\over p_1^0+m_1})\sqrt{{(p_{1'}^0+m_{1'})(p_1^0+m_1)\over p_{1'}^0
p_1^0}}
\end{eqnarray}
and
\begin{eqnarray}
h_V &=& {i\over m_I-m_{X^*}}\int d^3p_1\phi_X^*(\vec p_{1'})
\phi_I(\vec p_1)({p_{1'}^3\over p^0_{1'}+m_{1'}}
-{p_1^3\over p_1^0+m_1})\sqrt{{(p_{1'}^0+m_{1'})
(p_1^0+m_1)\over p_{1'}^0p_1^0}},\\
h_{A_1} &=& {i\over m_I+m_{X^*}}\int d^3p_1\phi_X^*(\vec p_{1'})
\phi_I(\vec p_1)\times \nonumber \\
&& (1-{p_{1'}^3 p_1^3
\over (p_{1'}^0+m_{1'})(p_1^0+m_1)})\sqrt{{(p_{1'}^0+m_{1'})
(p_1^0+m_1)\over p_{1'}^0p_1^0}},\\
h_{A_2} &=& {2(m_I+m_{X^*})^2\over 3m_I^2+m_{X^*}^2}h_{A_1}-
{i 4m_Im_{X^*}\over (m_I-m_{X^*})(3m_I^2+m_{X^*}^2)}
\int d^3p_1\phi_X^*(\vec p_{1'})\phi_I(\vec p_1)
[{p_{1'}^3\over p^0_{1'}+m_{1'}} + \nonumber \\
&& {p_1^3\over p_1^0+m_1}]\sqrt{{(p_{1'}^0+m_{1'})
(p_1^0+m_1)\over p_{1'}^0p_1^0}}.
\end{eqnarray}
where the $\phi_{(X, X^*)}$ are wavefunctions of $\psi_{(1s, 2s)}$ in the
momentum space, i.e. the Fourier transformed  (\ref{wav})  and
(\ref{wav1}), in the expressions, $\vec p_1$ and $\vec p_{1'}$ denote the
3-momenta of the quarks  which take part in the reaction in the initial and
final mesons, while $m_{1},\;
m_{1'}$ are their masses respectively. $p^3$ and $p^0$ correspond to the
third and the zero-th components of the concerned 4-momenta.
In the helicity-coupling picture, all momenta of the mesons
are along $\hat z$, so 
$$p_I\equiv |\vec p_I|,\;\; p^3_{X(X^*)}\equiv \pm |\vec p_{X(X^*)}|,$$
but the quark momenta can be along any directions.
In the CM frame of the decaying meson $\vec p_I=0$ and $|\vec p_{X(X^*)}|
={m_I^2-m_{X(X^*)}^2\over 2m_I}$ as $q^2=0$, thus one has
$$p_1+ p_2\equiv p_I=(M, \vec 0),\;\;\;{\rm and} \;\;\; p_{1'}+p_{2'}
\equiv p_{X(X^*)}=(p^0_{X(X^*)}, 0,0, p^3_{X(X^*)}). $$

The resultant formulae look quite different from that given in
ref.\cite{Wirbel}, but as a matter of fact, as $|\vec p|\gg M$, they
coincide with each other.

Substituting all the information into eqs.(\ref{PP}) and (\ref{PV}), we
can have the final numerical results.\\

\noindent{\bf III. The numerical results}
\vspace{0.3cm}

In the whole calculations, only $\beta$ is a free parameter and one can
fix it by the energy-minimum condition
$${\partial E\over \partial\beta}={\partial<H>\over\beta}=0.$$
Then one obtains
\begin{eqnarray}
\label{alp}
\beta_{1s} &=& ({4\mu\over 3\sqrt{\pi}a^2})^{1/3} \\
\label{alp1}
\beta_{2s} &=& ({6\mu\over 7\sqrt{\pi}a^2})^{1/3},
\end{eqnarray}
where $\mu$ is the reduced mass and $a$ is an average radius of the meson.
There can be an uncertainty for $a$ and $\mu$, it does not affect our
qualitative conclusion even though indeed the numerical results can
be declined by a few ten percents. (see below).

Even though $f_D$ is not well measured yet, there are some reasonable
estimated values, so we take $f_D=0.15$ GeV and $f_B=0.125$ GeV \cite{Rosner}.
Numerically we use
$$f_{\pi}=0.132,\; f_{K}=0.161,\; f_{\rho}=0.212,\; f_{K^*}=0.221$$
in GeV.

By the well-measured value $\alpha_s(M_Z^2)=0.118$ \cite{Data}, we have
$\alpha_s(m_b=5\; GeV)=0.203$, $\alpha_s(m_c=1.5\; GeV)=0.265$, and
$$c_1^{(D)}=1.26,\;\; c_2^{(D)}=-0.51$$
$$c_1^{(B)}=1.10,\;\; c_2^{(B)}=-0.23.$$
Our result is fully consistent with ref.\cite{Cher} obtained in terms of
RGE.

It is also noted that since $m_c$ is not very large, one can expect,
the real values of $c_{1,2}^{(D)}$ may deviate from that predicted by
the perturbative QCD calculation, for example, it is claimed that
a set of $c_1^{(D)}=1.26\pm 0.04$ and $c_2^{(D)}=-0.51\pm 0.05$ can
fit data better. However, in below we will rely on the perturbative QCD
and use the values obtained by RGE.

The corresponding $a_1^{(B, D)}$ and $a_2^{(B,D)}$ would depend on $\xi$ of
eq.(\ref{fac}).

For the radially excited $\psi_{2s}$ states, we will take $M_D(2s)\approx 2.4$
GeV
and $M_{\pi}(2s)=1.0$ GeV, $M_K(2s)=1.4$ GeV. By eq.(\ref{alp}), we fix
$$\beta_B=0.5,\; \beta_D(1s)=0.45, \; \beta_D(2s)=0.39,\; \beta_{\pi}(1s)=
0.3,\; \beta_{\pi}(2s)=0.26, \; \beta_K(1s)=0.4, \; \beta_K(2s)=0.34$$
in GeV. All the parameters are obtained according to eq.(\ref{alp}) and
(\ref{alp1}).

Numerically, we have
\begin{equation}
{\Gamma_{D^+}\over\Gamma_{D^0}}=\left\{  \begin{array}{ll}
0.9,\;\;\;\;\;\;\;\; & \delta=0 \\
0.70,\;\;\;\;\;\;\;\; & \delta=-0.5, \\
0.56, & \delta=-1.
\end{array} \right.
\end{equation}
It seems that the $\delta=-1$ solution suits the data on D-decays better than
other $\delta$ values and this conclusion was also predicted by Stech et al.
a long while ago \cite{Stech}.

For the B-case, without considering the $\psi_{2s}$ excited state contribution,
we have
\begin{equation}
{\Gamma_{B^-}\over\Gamma_{B^0}}=\left\{  \begin{array}{ll}
1.28,\;\;\;\;\;\;\;\; & \delta=0, \\
0.90,\;\;\;\;\;\;\;\; & \delta=-0.5, \\
0.59, & \delta=-1,
\end{array} \right.
\end{equation}
If one looks at $\delta=-1$ which is consistent with that obtained in D-decays,
the ratio is close to 0.5 as expected (see the introduction). When
we take into account the contributions from the $\psi_{2s}$ excited states,
the whole result is modified as
\begin{equation}
{\Gamma_{B^-}\over\Gamma_{B^0}}=\left\{  \begin{array}{ll}
1.02\;\;\;\;\;\;\;\; & \delta=0 \\
0.99\;\;\;\;\;\;\;\; & \delta=-0.5 \\
0.98 & \delta=-1
\end{array} \right.
\end{equation}
this result is very consistent with the data on the lifetimes of both
D and B-mesons. We will discuss this result in next section.\\

\noindent{\bf IV. Conclusion and discussion}

\vspace{0.3cm}

B and D mesons all contain a heavy quark and a light one, we have every
reason to believe that they have similar characteristics. Indeed a symmetry
between b and c quarks (B and D mesons) \cite{Isgur} is confirmed
by phenomenology. However, one obvious discrepancy that $\tau_{D^{\pm}}
\sim 2\tau_{D^0}$ while $\tau_{B^{\pm}}\sim\tau_{B^0}$ implies some
distinction between B and D mesons.

There have been alternative ways to interpret the lifetime difference 
of B and D. For example, Bander, Silverman and Soni \cite{Bander}
suggested that the reaction $D^0\rightarrow s+\bar d+\, gluon$ as a source
for the difference in the lifetimes of $D^0$ and $D^{\pm}$ and in other
way,
one can suppose that the factorization factor
$\delta$ can be different for B and D or the signs of $a_2$ can change etc.
However, if we consider similarities between B and D, it is natural to
accept an assumption that $\delta$ would not be too declined in B and D
cases. In literature \cite{Stech}, in D-physics, $\delta$ is very close to
$-1$ and our results confirm this allegation. Cheng found \cite{Cheng1}
that $r_2=-0.67, \;-(0.9-1.1)$ for $D\rightarrow \bar K\pi,\; \bar K^*\pi$
respectively, where our $\delta=(N_c/2)r_2$, it indicates that $\delta\sim
-1$. But to fit B-decay data, Cheng concluded $r_2=+0.36$ which drastically
deviates from the parameter for D-decays, so one would ask how it could  be so?

Instead, we accept the assumption that a symmetry between b and c holds and
$c_1^{(D,B)}$, $c_2^{(D,B)}$ can be derived with the RGE. Meanwhile we also
notice that since B-mesons are much heavier than D-mesons, there can be
radially excited states $\psi_{2s}^D$ and $\psi_{2s}^{\pi}$ as decay
products in B-decays, but not for D-decays. The $\psi_{2s}$ states may
cause the hadronic matrix elements to be in opposite sign to the
$\psi_{1s}$ final states and it would result in a change to make
$\tau_{B^{\pm}}\sim\tau_{B^0}$. Obviously, it is determined by an overlapping
integral between wavefunctions of the final and initial mesons. Our numerical
results show that the integrals for $\psi_{2s}$ and $\psi_{1s}$ can have
opposite signs depending on the parameter $\beta$. Our $\beta-$values are
reasonably determined by data, even though not very accurate. We show that
as $\delta\sim -1$, as taking into account the contribution from $\psi_{2s}
^{D,\pi}$ as well as $\psi_{1s}^{D,\pi}$, approximately
$$\tau_{B^{\pm}}\sim\tau_{B^0},\;\; \tau_{D^{\pm}}\sim 2\tau_{D^0}.$$

Our mechanism is in parallel to the PI effects discussed by some authors
\cite{Buras}\cite{bigi}. It
is based on the common knowledge that as long as all the exclusive channels 
(in fact, the main ones) are
summed up, the total width should be obtained, i.e. equivalent to the 
inclusive evaluation. Thus in our picture an interference between the decay 
products
of the $b$ ($c$) quark and the light one is automatically considered via the
$a_1$ and $a_2$ interference.

Since, indeed, we only consider the most Cabibbo favorable channels to
estimate the lifetimes, there can be contributions from the rare decays and
the numerical results can deviate a bit, but in general the same mechanism
proposed by Close and Lipkin can apply.
Hence the rule is the same
to all channels, namely $\psi_{2s}$ always contributes as
well as $\psi_{1s}$, our
results seem sufficiently convincing. As a matter of fact, the $\psi_{2s}$
is still light enough and there is large phase space available for B, but on
contrary, not for D-meson.

For evaluating the hadronic matrix elements, we use the non-relativistic
quark model. Even though the model is approximate, our qualitative
conclusion does not change.

Surely, we can make the ratios of lifetimes for D and B mesons perfectly
coincide with the data by carefully adjusting the $\beta$ values in
the wavefunctions. However, since there are many uncertain factors such as
the contributions of the rare decays, non-relativistic form of the
wavefunctions and the factorization factor $\delta$ etc, which make a very
accurate evaluation impossible, so only adjusting $\beta$ value to fit
data seems not necessary. In fact, as the most important point,
one can draw a qualitative conclusion confidently that the the contribution
of $\psi_{2s}$ is important to B-decays, namely the puzzle of the lifetimes
of B and D mesons can be reasonably explained away by its participation.

It is important to notice that not only the lifetimes of B-meson is in
contrary to our knowledge based on the perturbative QCD and D-physics if
the $\psi_{2s}$ contributions is not taken into account, but also
similar puzzles exist at
many channels of B-meson decays. It is that the value of $a_2$
is not universal \cite{a2} and
its sign is also uncertain. It is hard to understand. So we hope that
by taking into account of the $\psi_{2s}$ contributions, all the discrepancies
may get a reasonable explanation. Because the relatively heavy $\psi_{2s}$ is
still light to B-meson and does not affect its phase space integration
very much, so
maybe in measurements of exclusive channels, certain $\psi_{2s}$
with the same quantum numbers as $\psi_{1s}$ gets
mixed in and is not well tagged out. It causes the superficial discrepancy.
To carefully and thoroughly investigate the influence and effects
of possible $\psi_{2s}$ decay products in B-decays is
the goal of our next works. \\

\noindent {\bf Acknowledgment}

One of us (Li) would like to thank the Rutherford Appleton Laboratory
for its hospitality during his sabbatical visit and this work was
first originated over there. He is also indebted to Prof. F.E.Close
and L. Oliver for helpful discussions. This work is partly supported by 
the National Natural Science Foundation of China (NSFC)

\vspace{2cm}

\centerline{\bf Figure Captions}

Fig.1(a)-(d). The quark diagrams for the non-leptonic decays of B and D mesons
(here we take $D \rightarrow K \pi$ as an example).\\

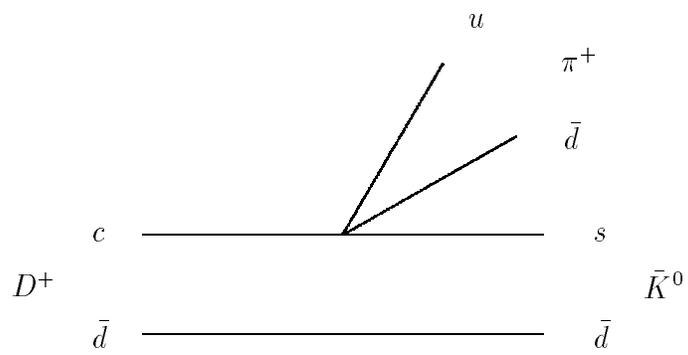
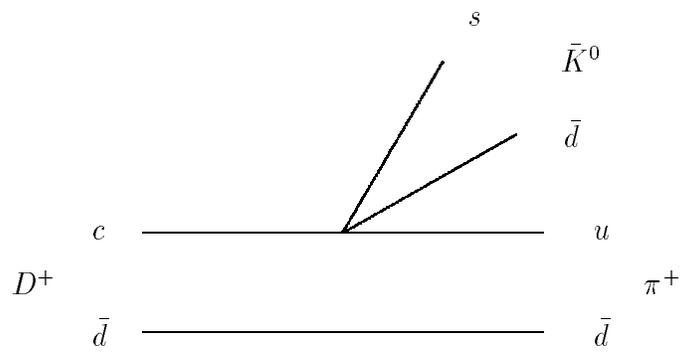
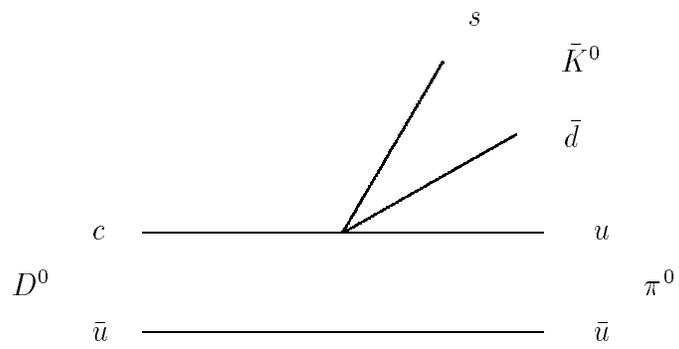
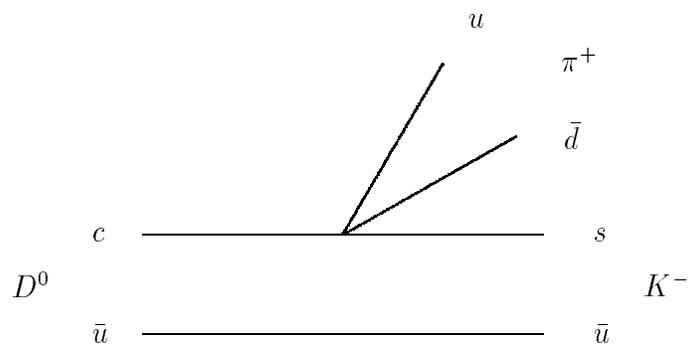
\end{document}